# Universal Phase Contrast in Micro-CT Systems

G. Lioliou[1*], A. Astolfo[1], M. Endrizzi[1], D. Bate[2,1], C. K. Hagen[1#] and A. Olivo[1#]
[1]Department of Medical Physics and Biomedical Engineering, University College London, UK
[2] Nikon X-Tek Systems Ltd, Herts, Tring, HP23 4JX, United Kingdom
* Corresponding author: G.Lioliou@ucl.ac.uk
# These two authors jointly supervised this work

Conventional high-resolution micro-CT systems are regarded as attenuation-based unless visible Fresnel fringes reveal the presence of propagation-based phase contrast. Here we show that this interpretation is incomplete. When propagation-induced phase transfer is non-negligible relative to the system blur, micro-CT operates in a phase-transfer regime in which propagation improves spatial resolution relative to that expected from source and detector blur alone, irrespective of whether Fresnel fringes remain visible. Because micro-CT systems possess finite source and detector blur, this phase transfer is accompanied by a degree of hardware-induced phase retrieval (HIPR), ranging from under-HIPR to matched-HIPR and over-HIPR. We further identify detector-induced phase retrieval (DIPR) as the optimal case in which the detector provides all of the filtering required for phase retrieval, thereby preserving the phase-induced resolution enhancement while introducing spatial correlations between detected photons that reduce high-spatial-frequency noise. We derive analytical conditions for HIPR and DIPR, introduce frequency-domain metrics quantifying preserved phase transfer and detector compatibility, and validate the framework numerically with simulations and experimentally using custom and commercial micro-CT systems. Our results demonstrate that phase-induced resolution enhancement can occur in conventional micro-CT systems even in the absence of visible Fresnel fringes, changing how image formation, spatial resolution, and optimization should be interpreted.

The use of micro-CT has seen an increase across a diverse range of industries, including medical research[1], materials science[2], and geology[3]. The continuous improvements in technology are driving a market expansion: in 2023, this was valued at USD 282.4 million and is projected to grow at a compound annual growth rate of about 6.7% through 2030[4].

Advances in micro-CT technology include the ability to capture images at high spatial resolution. In medical research, this has enabled quantification of lung volume, aerated tissue and lesion progression in disease models such as fibrosis, infection, or cancer[5], reconstruction of pulmonary microvasculature down to 6 μm[6,7], as well as detailed imaging of the heart[8] and brain microvasculature[9,10]. In materials science, high-resolution X-ray micro-CT enables non-destructive three-dimensional characterization of microstructure and its evolution, including electrode morphology and degradation in batteries, as well as porosity, defects and other microstructural features in additive manufacturing[11,12]. In industrial applications, micro-CT is increasingly used for non-destructive defect analysis, dimensional metrology and quality control of complex manufactured components, including internal features that are inaccessible to conventional surface-based inspection methods[13]. High spatial resolution in micro-CT is typically achieved using microfocus sources, since a small X-ray focal spot reduces the geometric blur[10]. The detector pixel size and/or its Point Spread Function (PSF), which are constrained by current detector technology[14,15], contribute additional blur. However, employing a large geometric magnification leads to a reduced impact of the blur arising from the detector, and thus higher spatial resolution. Such geometries position the sample close to the source resulting in non-negligible sample-to-detector distances.

Another technological advance is X-ray phase contrast CT (XPC-CT). XPC-CT offers superior contrast-to-noise ratio over conventional attenuation-based CT at equal dose, particularly for low-Z materials such as soft tissues[16], carbon fibres, and light polymers[17], by capturing phase shifts arising from variations in the real part of the complex refractive index[18] $\delta$ (while the imaginary part $\beta$ drives absorption). Since phase shifts cannot be measured directly, phase contrast techniques translate them into detectable intensity variations. Approaches include crystal interferometry[19], analyzer-based imaging[20], grating-based interferometry[21], propagation-based imaging (PBI)[22], and tracking approaches (using absorption grids[23], phase gratings[24], or speckle patterns[25]). Except for PBI, all require optical components.

PBI relies on an increased sample-to-detector distance, allowing X-ray beam propagation and formation of interference fringes through Fresnel diffraction. In micro-CT systems, it requires sufficient spatial coherence typically provided by microfocus sources[26] and high-resolution detectors to resolve the phase fringes, since both the source and detector influence the phase-contrast formation[27]. The free-space propagation interference fringes in the acquired images are caused by phase gradients, and lead to a measured intensity not directly linked to the sample's attenuation properties[28]. Quantitative XPC-CT requires the interference fringes to be converted into area contrast via the use of a phase retrieval method[29]. The most successful and widespread approach is Paganin's method[30,31]. It relies on low-pass frequency-domain filtering with an optimised regularization parameter, which is consistent with the near-field diffraction physics of phase-contrast imaging. It is specifically designed to preserve fine structural details while suppressing noise and fringes[31]. Increases of signal-to-noise ratio (SNR) by up to two orders of magnitude over conventional CT at the same radiation dose, without loss of image quality, have been reported[32]. Paganin retrieval enables the recovery of the phase of the sample from the registered intensity distribution (acquired image), using the homogeneous variant of the Transport of Intensity Equation (TIE). The image formation physics underlying propagation-based imaging, including the interplay between free-space propagation, finite source size, detector blur and phase retrieval, have been extensively studied by Gureyev and co-workers[28-39]. Their work describes theoretical relationships between propagation-induced edge enhancement, detector response and partial phase retrieval. The present analysis builds on their work, in particular their *deblur by defocus* concept[35], by providing a new interpretation in the context of micro-CT, leading to new understanding of image formation, spatial resolution and optimisation of such systems.

Considering that high-resolution micro-CT systems typically employ a small X-ray focal spot together with an extended sample-to-detector distance, propagation-induced phase transfer can contribute to image formation even when no visible Fresnel fringes are observed. In practical systems, the finite source and detector responses partially (or fully) counteract the propagation-induced sharpening, resulting in a degree of hardware-induced phase retrieval. Consequently, the absence of visible fringes does not imply purely absorption-based image formation: propagation may still modify the effective system response and improve spatial resolution relative to that expected from the conventional source and detector blur alone. Neglecting phase effects may therefore lead to an incorrect interpretation of spatial resolution in conventional micro-CT images.

Here we establish a framework that distinguishes attenuation-only imaging from phase-transfer imaging in micro-CT according to the relative contributions of propagation-induced sharpening and system blur. Within the phase-transfer regime, the imaging hardware may provide under-, matched-, or over-retrieval. We identify detector-induced phase retrieval (DIPR) as the optimal case in which the detector provides all of the filtering required to compensate the propagation-induced phase transfer, thereby retaining the phase-induced resolution enhancement while also providing the noise-suppression benefit of detector-based filtering. We derive the analytical conditions governing phase-transfer imaging and DIPR, introduce two frequency-domain metrics quantifying preserved phase transfer and detector compatibility, and validate the framework numerically and experimentally using a custom-built and a commercial micro-CT scanner. The resulting framework provides a unified basis for identifying, interpreting and optimizing propagation-induced phase effects in custom and commercial micro-CT systems.

## Results

### Theoretical prediction of phase-transfer imaging and DIPR

We first evaluate the phase-transfer and hardware-retrieval conditions for the investigated geometries of our custom system and the Nikon scanner. As is described in Methods, an important distinction should be made between the effects of source blur and detector blur. Both suppress the propagation-induced edge enhancement and therefore contribute to hardware-induced phase retrieval (HIPR). However, they do so through different physical mechanisms, which have different implications for image quality and for the optimization of laboratory micro-CT systems. We therefore first assess whether sufficient propagation-induced phase transfer survives finite-source blurring to enable DIPR. For all geometries considered, the propagation distance is non-zero and the finite-source coherence criterion (see Methods, Eq. (9)), is fulfilled. Figure 1 (a) shows the residual propagation-induced sharpening after accounting for the projected finite-source blur, which remains positive throughout the investigated range for both systems. Thus, source blur does not fully compensate the propagation-induced sharpening before detection, leaving phase information available for subsequent detector-based retrieval. Increasing the source size reduces this residual phase transfer and narrows the range of geometries satisfying the finite-source criterion, as illustrated for the large-focus mode of our custom system in Supplementary Fig. 1.

We next quantify the magnitude of the phase-to-intensity transfer using the metric $n_\varphi$ (see Methods, Eq. (13)), which computes the fraction of the coherent-limit phase transfer retained after accounting for finite-source blur over the usable spatial-frequency range of each imaging system. This frequency range is defined as the freqiencies below the onset of spectral overlap (see Methods). Figure 1 (b) shows $n_\varphi$ as a function of source-to-sample distance, $R_1$, for the two imaging systems investigated. For a coherent point source, this metric varies with $R_1$ in accordance with the effective propagation distance, $R'$, variation with $R_1$; it increases as $R_1$ increases, it reaches its maximum when the effective propagation distance is maximized at $R_{tot}/2$, and it then decreases for further $R_1$ increases. For finite source sizes, however, the projected source blur decreases as $R_1$ increases, partially compensating for the reduction in effective propagation distance. Consequently, the maximum of $n_\varphi$ shifts to $R_1$ values slightly larger than $R_{tot}/2$, with the shift increasing for larger source sizes. This occurred at 485 mm for our system, where $n_\varphi = 0.969$, and 460 mm for the Nikon scanner, where $n_\varphi = 0.996$. A finite source size also reduces the maximum attainable value of $n_\varphi$ relative to the coherent limit (where $n_\varphi = 1$). However, it should be noted that $n_\varphi$ is evaluated over the usable spatial-frequency range of each imaging system. The Nikon scanner, with its larger effective detector pixels, has a narrower usable spatial-frequency range at a given magnification. In contrast, our custom system has smaller detector pixels and a broader usable spatial-frequency range, so $n_\varphi$ is evaluated over a wider range of spatial frequencies, including higher frequencies that are more strongly affected by a finite source blur. This explains the larger $n_\varphi$ value achieved with the Nikon scanner compared to our custom system. Therefore, $n_\varphi$ captures the interplay between propagation distance and partial source coherence. The metric $n_\varphi$ is calculated for other imaging systems and the results are presented in Supplementary Fig. 2, for comparison purposes.

Figure 1 (c) presents the detector compatibility metric $E_{match}$, Eq. (14), which quantifies the mismatch between the detector MTF and the Paganin filter over the usable spatial-frequency range. For both imaging systems, $E_{match}$ exhibits a well-defined minimum, identifying the geometry at which the detector most closely reproduces the required Paganin low-pass filtering. A value of $E_{match} = 0.03$ was calculated for our system and the Nikon scanner at 465 mm and at 150 mm, respectively. Importantly, $E_{match} \leq 0.04$ over the ranges 430 mm – 500 mm for our custom system and over 130 mm – 170 mm for the Nikon scanner, indicating that near-matched DIPR can be achieved over a finite range of magnification values rather than a single one. At smaller $R_1$ values, the detector provides less filtering than required by the Paganin transfer function (i.e., for optimal phase retrieval), corresponding to detector under-retrieval and leaving residual propagation-induced edge enhancement. Conversely, at larger $R_1$, the detector provides greater low-pass filtering than required, corresponding to detector over-retrieval and increased reduction of high spatial frequencies. The minimum of $E_{match}$ defines the optimum DIPR geometry. It should be noted that, due to the detectors' MTF having a different shape than the Paganin filter, an exact spectral match ($E_{match} = 0$) cannot be achieved.

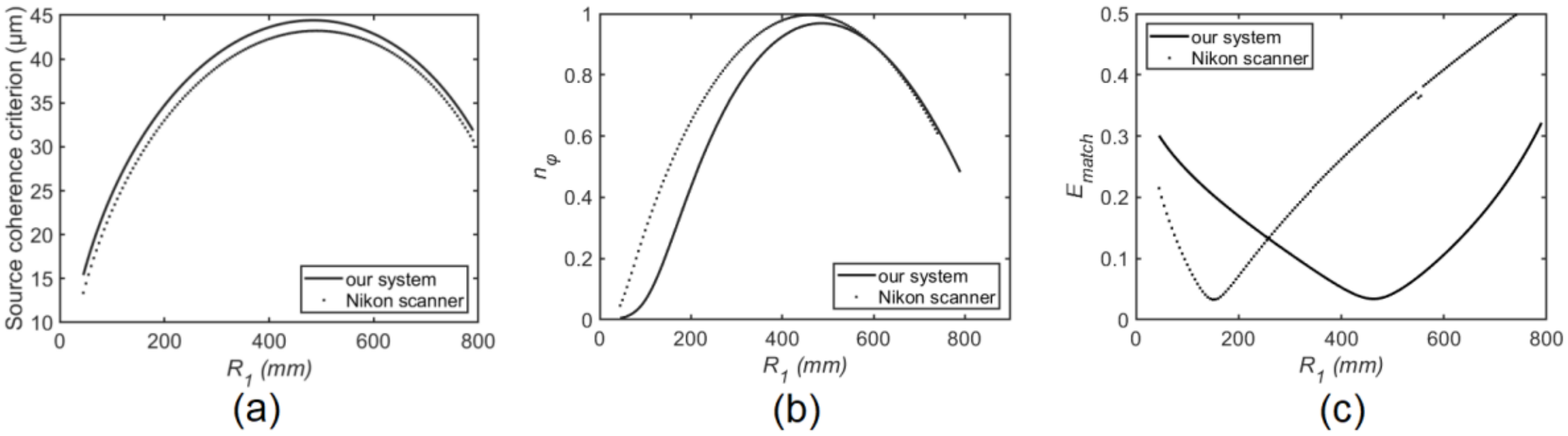

Figure 1. (a) Finite-source coherence criterion, (b) normalized pre-detector phase-transfer metric, $n_\phi$, and (c) detector compatibility metric, $E_{match}$, as functions of source-to-sample distance, $R_1$, for our system (black solid line) and the Nikon scanner (black dots).

Finally, the conventional system blur and its implications for spatial resolution are evaluated. The FWHM corresponding to the equivalent Gaussian system width, $\sigma_{sys}$, comprising the source and detector contributions, Eq. (16), is shown in Fig. 2. It increases from 14.5 µm at $R_1$ = 45 mm to 157.7 µm at $R_1$ = 790 mm for our custom system. Similarly, it increases from 24.5 µm at $R_1$ = 45 mm to 322.5 µm at $R_1$ = 790 mm for the Nikon scanner. For both systems, the detector contribution dominates the total system blur. Because the projected detector blur increases with decreasing magnification, the equivalent system blur increases monotonically with increasing source-to-sample distance $R_1$. Compared with our custom system, the Nikon scanner exhibits a larger equivalent system blur throughout the investigated range due to its broader detector PSF. For our custom system (Fig. 2 (a)), the total system blur exceeds the sampling limit over the entire range of investigated geometries. Consequently, the phase-induced reduction in effective system width predicted by Eq. (6) can directly translate into improved measured spatial resolution. In contrast, for the commercial Nikon scanner (Fig. 2 (b)), the predicted total system blur falls below the sampling limit for $R_1$ > 95 mm. At this position, the system is sampling-limited even without consideration of phase effects, such that the effective spatial resolution is therefore expected to become sampling-limited rather than being determined by the continuous system blur.

At the optimum DIPR geometry, which is sample dependent, the intrinsic spatial resolution predicted from Eq. (10), corresponding to the theoretical optimum spatial resolution (similarly, sample dependent), is 5.9 µm FWHM for our custom system and 14.2 µm FWHM for the Nikon scanner, compared with equivalent total system blurs of 93.0 µm and 62.8 μm FWHM, respectively. This improvement arises from propagation-induced phase transfer, which approximately compensates the detector blur, leaving the projected source blur as the dominant limiting factor. For the investigated geometries, the corresponding sampling-limited resolutions are 50.6 μm and 65.9 μm for our custom system and the Nikon scanner, respectively. Therefore, although DIPR substantially improves the intrinsic system resolution, the full benefit can only be realized when the sampling is sufficiently fine to capture the available high-spatial-frequency information. To overcome this sampling limitation, so-called sub-pixel dithering, referring to performing identical CT acquisitions with a relative effective sub-pixel sample translation, can be employed, reducing the effective sampling pitch without altering the detector pixel size (technically "aperture") or the pre-detector phase-to-intensity transfer. For the Nikon scanner and for a dithering step equal to one-half of the effective pixel size at $R_1$ = 150 mm, the sampling-limited spatial resolution improves from 65.9 μm to 32.9 μm. Although this remains above the intrinsic DIPR-limited resolution, it suggests that dithering can enable the realisation of at least a degree of phase-induced resolution enhancement. The predicted spatial resolution of our custom system with the large focus mode is presented in Supplementary Fig. 3.

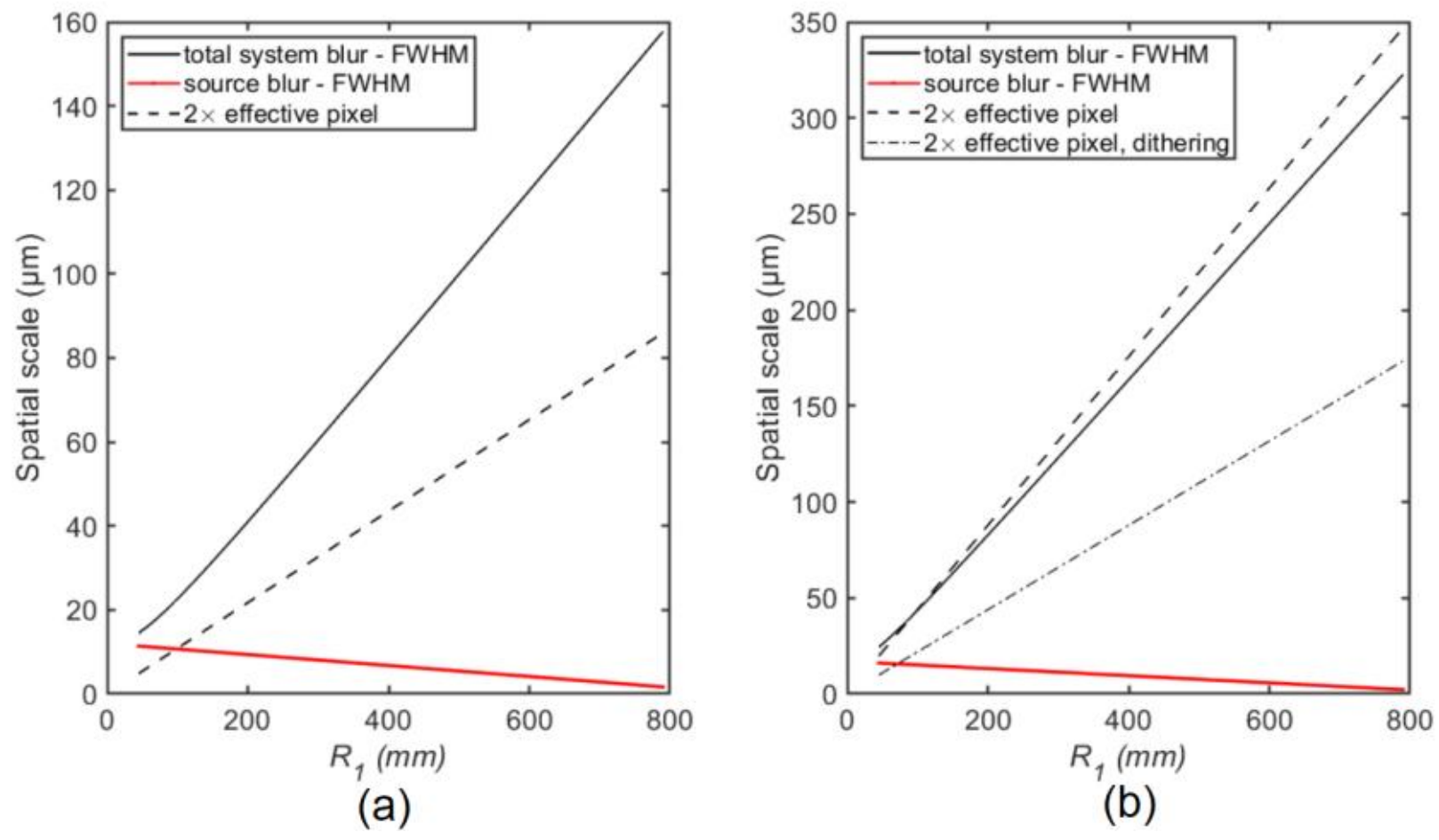

Figure 2. Total system blur (black solid line), source blur (red solid line), and 2x effective pixel without (black dashed line) dithering as functions of source-to-sample distance, $R_1$, for our (a) custom system and (b) the Nikon scanner (also including 2x effective pixel with (black dash-dotted line) dithering).

**Experimental demonstration of phase-transfer imaging and DIPR in our custom system**

Here, we experimentally demonstrate the transition from under-HIPR through matched-HIPR, (specifically, DIPR to over-HIPR), by varying the source-to-sample distance while maintaining a fixed source-to-detector distance.

Figure 3 (d) shows an axial reconstructed CT slice of the biological sample (a tissue-engineered esophageal scaffold, see Methods) for each of the following 4 scans. Scan 1 corresponds to strongly under-HIPR propagation-based phase imaging, representing the conventional edge-enhanced PBI condition. Its unretrieved reconstructed slice has poor soft tissue SNR and as expected, shows edge enhancement arising from the phase fringes. Based on previous work[40], the $\delta/\beta$ of this sample is ~1000 at the mean energy of the source operated at 60 kVp. The Paganin filter was applied to the projection images of Scan 1, with the corresponding reconstructed slice shown in Fig. 3 (b).

Scan 2 – Scan 4 verify the predicted transition towards, close to, and beyond the DIPR regime; the detector compatibility metric $E_{match}$ is 0.10, 0.04, and 0.11, respectively. This indicates an under-retrieved geometry (Scan 2), a geometry close to the optimum DIPR condition (Scan 3), and an over-retrieved geometry (Scan 4). This progression is evident in Fig. 3(c), where the detector MTF evolves from providing insufficient low-pass filtering relative to the Paganin filter (Scan 2), to a near spectral match (Scan 3), and finally to excessive low-pass filtering (Scan 4).

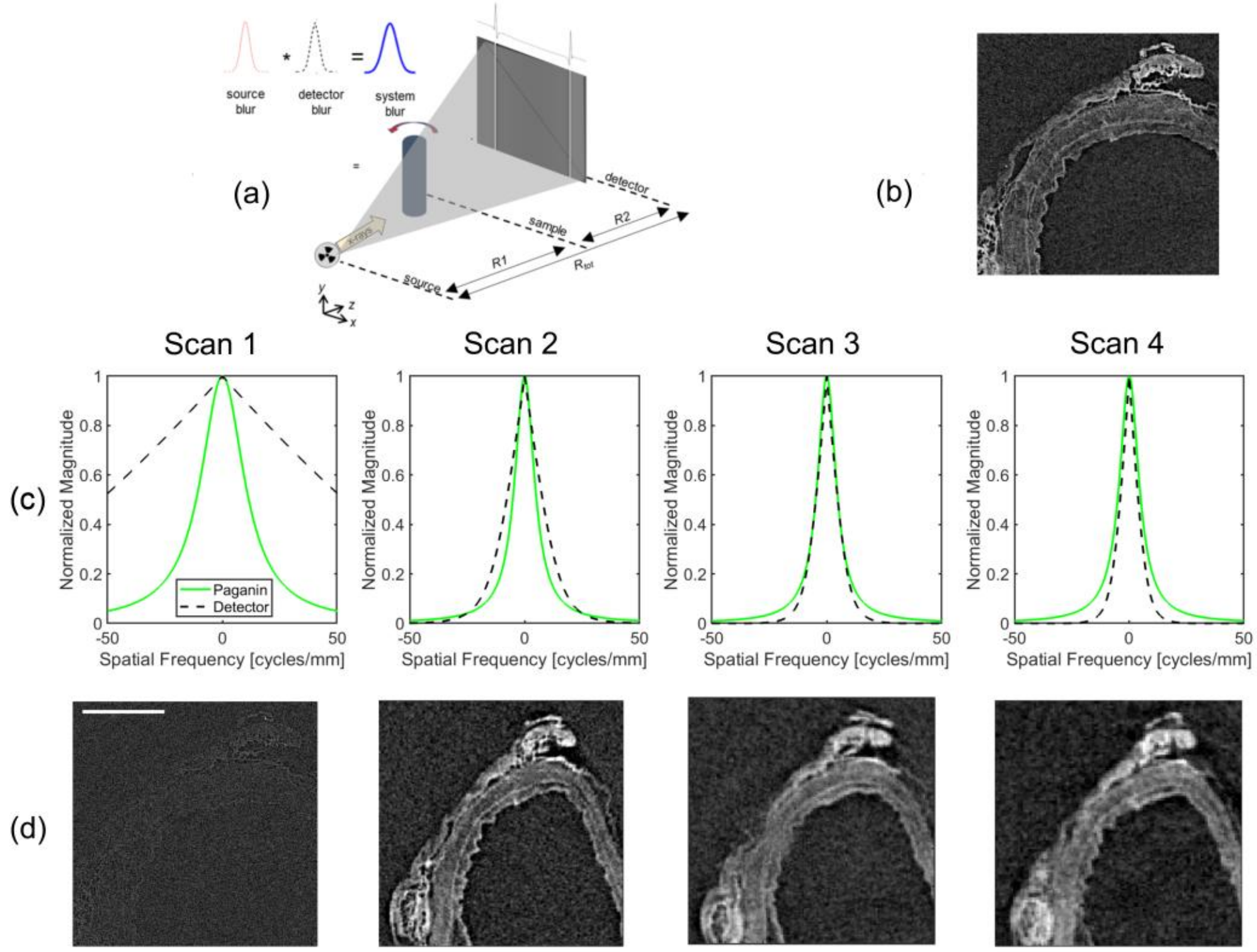

Figure 3. (a) Schematic diagram of our custom system and associated system blur, (b) Paganin retrieved reconstructed slice of the biological sample, from Scan 1. For each of the four scans, Scan 1 – Scan 4, we show (c) the detector blur (black dashed line) the Paganin filter (green solid line) in the frequency domain, and (d) the reconstructed images of the same biological sample with no additional (software) retrieval. Note that magnification (and image pixel size) differs between scans, and that the biological sample slightly deformed between scans.

The experimentally acquired images of the 1 mm diameter PTFE wire (see Methods) were used for quantitative analysis and to validate the simulation framework. A comparison between the experimental and simulated normalized projections along the PTFE wire for Scan 1 – Scan 4 is presented in Fig. 4 (a). The first observation is that, as the detector contribution to hardware retrieval increases, the relative fringe amplitude at the PTFE edges decreases: it reduces from 0.05 ± 0.02 for Scan 1, to 0.013 ± 0.003 for Scan 2, and becomes indistinguishable from the noise floor for Scan 3 and Scan 4. The experimental relative fringe amplitude agrees with the simulated one within uncertainty: 0.04 for Scan 1, 0.016 for Scan 2, while no residual fringes are observed for Scans 3 and 4 in simulation. Figures 4 (b) and (c) show the experimental and simulated reconstructed images of the PTFE wire for Scan 1 – Scan 4, with no additional (software) retrieval applied. In addition to the agreement between experiment and simulation, the different SNR values of the experimental reconstructed images from different scans can be appreciated. The SNR is 2.3, 34.5, 64.4, 69.6, for Scan 1 – Scan 4, respectively. For comparison purposes, the Paganin retrieved reconstructed image of Scan 1 is shown in Fig. 4 (d); this has an SNR of 76.7. The further SNR increase in Scan 4 is a consequence of over-HIPR and must therefore be considered together with the associated loss of spatial bandwidth. All four unretrieved reconstructions are sampling-limited at their respective acquisition geometries, corresponding to two-pixel spatial scales of 4.9 μm (2 pixels), 33.1 μm (2 pixels), 54.4 μm (2 pixels) and 65.8 μm (2 pixels), respectively. For comparison purposes, the Paganin retrieved reconstructed image has a spatial resolution of 14.7 μm (6 pixels), in accordance with the estimated value (Fig. 2 (a)), of 14.5 μm. An experimental reconstructed image for another example scan is shown in Supplementary Fig. 4. A comparison of the reconstructed profiles along the PTFE for Scan 1 – Scan 4 is shown in Supplementary Fig. 5.

This comparison demonstrates that hardware phase retrieval changes the interpretation of image formation, noise and spatial resolution when phase-transfer is present. Scan 3 closely approximates the corresponding Paganin retrieved image, with the retrieval obtained at the point of acquisition. For a fixed source-to-detector distance, increasing the source-to-sample distance $R_1$ reduces the geometric magnification and therefore increases the effective sample-plane pixel size. Assuming photon-noise-limited imaging, the number of detected photons per sample pixel increases proportionally to the pixel area, resulting in a corresponding improvement in SNR. Comparing Scan 1 and Scan 3, and considering their different magnification and exposure time, a 22× increase in SNR is expected for Scan 3 c.f., Scan 1, from 2.3 to 50.6. Under the DIPR condition, the detector additionally performs low-pass filtering, further suppressing high-frequency noise and reducing the image variance, here by a further 27%. Consequently, the experimentally observed SNR is determined by the combined effects of photon statistics and detector-induced filtering.

The spatial resolution is similarly modified by phase transfer. Under DIPR, propagation-induced sharpening approximately compensates the detector blur. The remaining intrinsic blur is therefore primarily governed by the projected source size, while the experimentally achievable resolution here is ultimately limited by detector sampling. An improvement from 100 µm to 54.4 µm FWHM spatial resolution is achieved at 500 mm $R_1$. The DIPR regime thus provides 1) phase contrast directly at acquisition and 2) a spatial resolution better than that predicted from the conventional system blur alone.

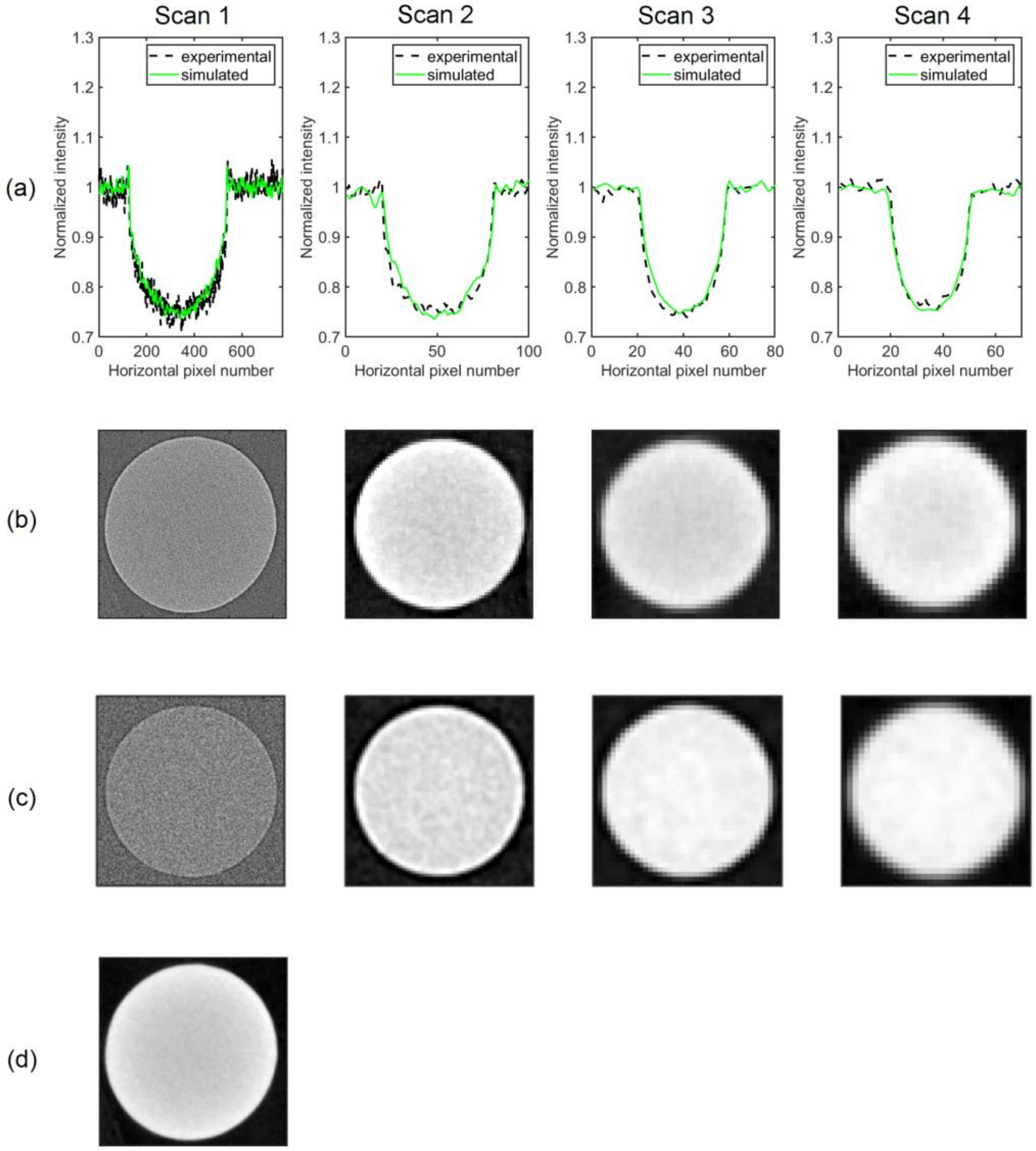

Figure 4. For each of the four scans, Scan 1- Scan 4, we show (a) experimental (black dashed line) and simulated (green solid line) profiles from normalized projections along the PTFE wire, (b) the experimental and (c) simulated reconstructed images with no software retrieval, and (d) the software (Paganin) retrieved reconstructed image from experimental Scan 1.

To further validate the theoretical and simulation framework, the relative fringe amplitude extracted from planar projection images over a range of $R_1$ values is compared with numerical simulations; this is shown in Fig. 5. Agreement is observed over the entire investigated range, confirming that the model predicts the evolution of propagation-induced phase contrast with the varying imaging geometry. At small $R_1$, corresponding to under-HIPR (conventional edge-enhanced PBI), Fresnel fringes are present and software phase retrieval (Paganin filter) is required. As $R_1$ increases, the fringe amplitude decreases monotonically due to the increasing detector-induced filtering, approaching zero near the predicted DIPR condition. Importantly, the disappearance of visible fringes does not indicate the absence of phase-to-intensity transfer; rather, it demonstrates that the detector intrinsically performs a low-pass filtering comparable to that of Paganin retrieval. Beyond the optimum DIPR region, over-HIPR, the fringe amplitude remains negligible while the detector increasingly over-retrieves the image, resulting in progressive suppression of high-spatial-frequency information. These results provide direct experimental evidence for the continuous transition from under-HIPR, to matched-HIPR (here DIPR, specifically), to over-HIPR.

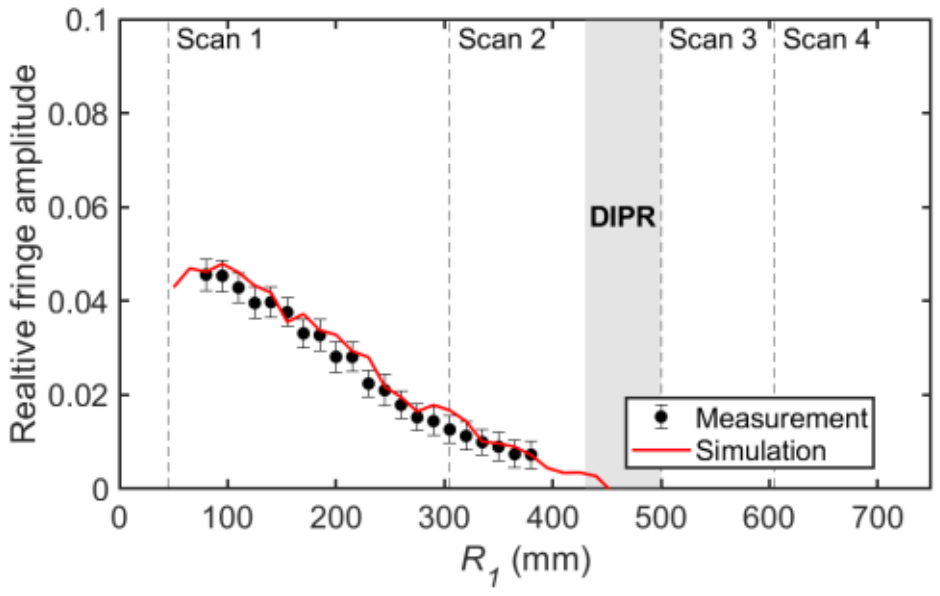

Figure 5. Experimentally measured (black circles) and simulated (red solid line) relative fringe amplitude as a function of $R_1$. Vertical dashed lines indicate the four experimental scan positions, while the shaded region identifies the predicted near-matched DIPR range.

**Experimental demonstration of DIPR in a commercial (Nikon) micro-CT scanner**

Having experimentally validated the framework using our custom system, we next investigate its implications for a commercial micro-CT scanner. Here we examine whether DIPR occurs under realistic commercial micro-CT imaging conditions and its implications. The theoretical analysis predicts near-matched DIPR for source-to-sample distances between 130 and 170 mm, where the detector MTF closely approximates the required Paganin filtering. This provides an opportunity to test whether phase-induced resolution enhancement and hardware retrieval can occur under routine commercial micro-CT acquisition conditions despite the absence of visible Fresnel fringes.

We first consider the standard acquisition without dithering (Scan I), which represents the conventional operating mode of the commercial Nikon scanner. Based on our theoretical framework, this imaging geometry ($R_1$ = 130 mm) lies within the predicted near-matched DIPR range. The corresponding PTFE wire reconstruction is shown in Fig. 6 (a), together with the numerical simulation in Fig 6 (b). Close agreement is observed between experiment and simulation, with both exhibiting no measurable edge enhancement. The predicted equivalent total system blur at this geometry is 55.0 μm FWHM, while the source blur is 14.6 μm FWHM (Fig. 2 (b)); these correspond to the expected spatial resolution of Regime A and to the intrinsic resolution predicted in the matched-DIPR limit (sample dependent), respectively. The extracted spatial resolution of both the experimental and simulated reconstructed images of the PTFE wire of Scan I is 57.0 μm, corresponding to the sampling limit of 2x effective detector pixels. In this case, since both the conventional system-blur prediction and the DIPR intrinsic predictions are finer than the sampling-limited resolution, the reconstructed image alone cannot distinguish between conventional absorption imaging (Regime A) and HIPR (Regime B).

To overcome the sampling limitation and observe the predicted phase-induced improvement in spatial resolution in a commercial scanner, half-pixel dithering is performed by acquiring a second scan with a detector shift corresponding to half of the effective pixel size and interleaving the projection images prior to reconstruction. The effective sampling density is doubled, while the detector aperture remains unchanged. Unlike approaches that combine sub-pixel sampling with deconvolution to compensate for the detector PSF[41], however, dithering alone does not recover spatial frequencies rejected by the detector aperture. Consequently, it primarily benefits imaging systems that are sampling-limited, as is the case considered here. The resulting PTFE wire reconstruction is shown in Fig. 6 (a), together with the corresponding simulation in Fig. 6 (b), for Scan II. The measured resolution reaches the new sampling limit of 28.5 μm, demonstrating that the intrinsic system blur at this magnification is ≤ 28.5 μm, substantially finer than the 55.0 μm FWHM predicted from the conventional source/detector system blur. In other words, the conventional system blur cannot correctly describe the experimentally observed resolution. This provides direct experimental evidence of phase-induced resolution enhancement despite the absence of visible Fresnel fringes, consistent with the predicted near-matched DIPR condition for this geometry. A comparison of the reconstructed PTFE edge profiles with and without dithering is shown in Supplementary Fig. 6, confirming that half-pixel dithering lowers the sampling limit and reveals spatial-frequency information that is inaccessible in the standard acquisition. For comparison, Supplementary Fig. 7 (b) shows simulations performed with $\delta = 0$, corresponding to Regime A. The reconstructed absorption-only image exhibits a resolution consistent with the conventional total system blur, confirming that the experimentally observed improvement originates from DIPR.

Both the experimental and simulated dithered reconstructions exhibit a periodic texture. The reproduction of the same pattern in the numerical simulations indicates that it arises from the dithering and sampling process rather than from detector imperfections or experimental corrections. Half-pixel dithering doubles the sampling density while preserving the original detector aperture (200 μm), so adjacent samples in the interleaved projections originate from different detector integrations rather than from physically smaller detector elements. A comparison with simulations performed using a detector with a physical pixel size of 100 μm is provided in Supplementary Fig. 7 (c), for which the periodic texture is absent. This confirms that dithering increases sampling density but is not equivalent to reducing the physical detector aperture.

The reconstructed images of the biological sample with Scan I and Scan II are shown in Fig. 6 (c). Both reconstructions exhibit negligible visible edge enhancement, as expected for the predicted near-matched DIPR condition. Although half-pixel dithering provides access to higher spatial frequencies, the associated periodic texture partially obscures the improvement in the biological-sample reconstruction, such that the visual appearance remains comparable to that of the standard acquisition despite the quantitatively demonstrated improvement in sampling-limited resolution.

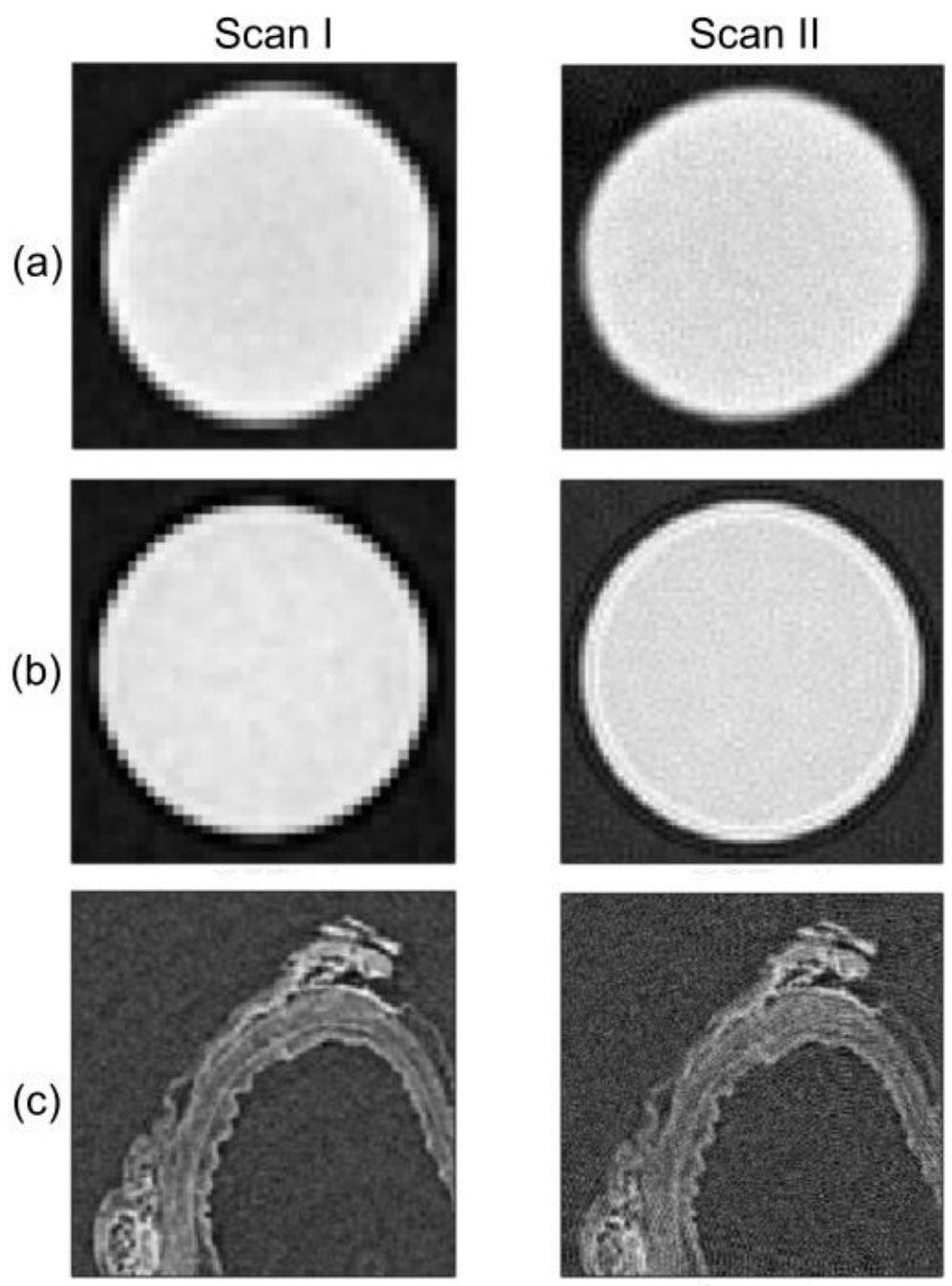

Figure 6. For each of the two scans, Scan I – Scan II, we show (a) the experimental and (b) simulated reconstructed images of the PTFE wire, and (c) the reconstructed images of the biological sample, all with no software retrieval.

# Discussion

The principal finding of this work is that propagation-induced phase transfer can contribute to image formation and improve spatial resolution in custom and commercial micro-CT systems even in the absence of visible Fresnel fringes. We show that micro-CT systems with non-negligible propagation-induced phase transfer operate in a phase-transfer regime (Regime B), in which propagation modifies the effective system response and can provide spatial resolution finer than that predicted from the conventional source and detector blur alone. Because practical imaging systems possess finite source and detector responses, this phase transfer is accompanied by HIPR, ranging from under-HIPR through matched-HIPR to over-HIPR. Importantly, suppression of the propagation-induced edge enhancement by the hardware response does not eliminate the underlying phase contribution to image formation. Consequently, commercial high-resolution micro-CT systems do not typically operate as purely absorption-based imaging systems despite producing images without visible Fresnel fringes, and the absence of such fringes should not be interpreted as evidence for the absence of propagation-induced phase effects.

Beyond providing new understanding on the image-formation mechanism, this framework changes the interpretation of detector blur in phase-transfer imaging. In Regime B, the detector response can contribute to hardware phase retrieval by low-pass filtering the propagation-induced phase transfer. Of particular interest is the DIPR regime (sample dependent), in which the detector MTF closely approximates the filtering required for Paganin retrieval. Under this condition, propagation-induced sharpening compensates to a good approximation the detector blur, while the detector response simultaneously suppresses high-spatial-frequency noise. The intrinsic spatial resolution can therefore be finer than that predicted from the conventional total system blur and is governed primarily by the residual source blur. Our experimental results demonstrate this effect directly: spatial resolution finer than that predicted from the conventional system blur was observed even when no visible Fresnel fringes were observable.

These findings have important implications for the design, optimization and use of custom and commercial micro-CT systems. Detector characteristics, source size, propagation geometry and detector sampling should be optimized jointly, as they collectively determine the balance between phase transfer, hardware retrieval and achievable spatial resolution. In particular, the source-to-sample distance becomes an additional design parameter that can be selected to achieve sufficient propagation-induced phase transfer while approaching matched hardware retrieval and avoiding excessive filtering in the over-HIPR regime. Where possible, the required retrieval filtering should be predominantly provided by the detector rather than by source blur, to retain the phase-induced resolution enhancement while maximizing the associated noise-suppression benefit. Importantly, the geometry that optimizes these phase-induced benefits does not generally coincide with that providing the smallest projected source blur or highest geometric magnification. Instead, it reflects a balance between propagation-induced phase transfer, source and detector contributions to hardware retrieval, and sampling. This optimum may occur at larger source-to-sample distances, enabling an increased FoV while maintaining spatial resolution substantially finer than that predicted from the conventional total system blur alone.

Collectively, these findings establish phase-transfer imaging as an important operating regime in custom and commercial micro-CT, in which propagation-induced phase effects can enhance spatial resolution even in the absence of visible Fresnel fringes. The finite source and detector responses determine the degree of HIPR, by providing the required low-pass filtering during acquisition. Within this framework, DIPR represents the optimized detector-dominated condition, in which the detector provides most or all of the required phase-retrieval filtering, combining phase-induced resolution enhancement with noise suppression. The analytical framework presented here provides practical criteria for identifying these conditions and establishes a basis for the joint optimization of propagation geometry, source and detector response, and sampling in custom and commercial micro-CT systems.

# Methods

## Theory

The theoretical framework presented here establishes (a) the conditions under which propagation-induced phase effects contribute to image formation, defining the transition from absorption imaging (Regime A) to phase transfer imaging (Regime B), (b) the conditions under which the imaging hardware provides under, matched or over "Paganin-like" retrieval, and (c) the resulting spatial resolution of the imaging system.

The following uses the general signal transmission theory[33]. i.e., Step 1 – signal encoding, Step 2 – transmission through the noisy channel, Step 3 – signal decoding. The information regarding the sample is imprinted on the complex X-ray wavefield traversing the sample, $U_0(\mathbf{r}, z = 0) = A(\mathbf{r})e^{i\phi(\mathbf{r})}$, where $A(\mathbf{r}) = \sqrt{I_0(\mathbf{r})}$ is the absorption term and $\phi(\mathbf{r})$ is the phase shift induced by the object, and $\mathbf{r}$ is the transverse spatial coordinate vector. The corresponding exit-surface intensity, $I_0(\mathbf{r}) = | U_0(\mathbf{r}) |^2$, is the signal recovered through Paganin retrieval.

Step 1 is the deterministic component of the imaging chain, comprising propagation-induced phase encoding, geometric blur arising from the finite X-ray source, and detector blur. The forward deterministic imaging model for the intensity (detected photon fluence) distribution after free-space propagation over distance $z = R_2$ for a monomorphous object, surrounded by air, is expressed as

$$I(r, R_2) \approx \left\{ \left(1 - a^2 \nabla_\perp{}^2\right) I_0(\mathbf{r}) \right\} * G\left(\mathbf{r}, \sigma_{sys}(M)\right), \tag{1}$$

where $I_0(\mathbf{r}) = I_{in} \exp[-B(\mathbf{r})]$, $I_{in}$ is the photon fluence of the incident beam, $\exp[-B(\mathbf{r})]$ is the X-ray transmission function of the object, $G(\mathbf{r}, \sigma_{sys}(M))$ is the PSF of the imaging system with a variance of $\sigma_{sys}^2$ (defined below), $\alpha^2 = \gamma R' \lambda/(4\pi)$ is the propagation parameter, $R' = R_2/M$ is the effective propagation distance, $M$ is the geometric magnification (defined below), $\gamma = \delta/\beta$, and $\nabla_\perp$ is the transverse Laplacian operator[34]. Here, $B(\mathbf{r}) = (4\pi/\lambda)\beta(\mathbf{r})T = \mu(\mathbf{r})T$, where $T$ is the projected thickness, and $\mu$ is the attenuation coefficient. It should be noted that Eq. (1) is the two-dimensional generalization of Eq. (5) in Gureyev et al.[34], obtained by replacing the one-dimensional second derivative with the transverse Laplacian. Equation (1) explicitly shows that free-space propagation converts curvature of phase into intensity modulation through a Laplacian operator. The Laplacian-type edge enhancement means that propagation sharpens interfaces and can reduce the apparent edge width relative to contact imaging[39].

In the Fourier domain, let $\hat{g}(\mathbf{f}) = \mathcal{F}\{g(\mathbf{r})\}$ denote the 2D Fourier transform of a generic function $g(\mathbf{r})$ (spatial frequencies $\mathbf{f}$ in cycles/m), then $\mathcal{F}\{\nabla_\perp^2 g\} = -(2\pi|\mathbf{f}|)^2\hat{g}(\mathbf{f})$. Taking the Fourier transform of Eq. (1), the normalized detected intensity over the incident intensity $I_{in}$ in Fourier domain including a generic system MTF, $H_{sys}(\mathbf{f}) = H_{det}(\mathbf{f})\, H_s(\mathbf{f})$, becomes

$$\widehat{I_{\text{det}}}(\mathbf{f}) = [1 + \alpha^2(2\pi \mid \mathbf{f} \mid)^2]\widehat{I_0}(\mathbf{f})H_{\text{sys}}(\mathbf{f}). \quad (2)$$

Equation (2) shows that the detected spectrum is the product of three factors: the normalized exit-surface intensity spectrum, $\widehat{I_0}(\mathbf{f})$, the forward TIE transfer function, and the system (source and detector) MTF, $H_{sys}(\mathbf{f})$. The propagation transfer function increases the relative weight of higher spatial frequencies, enhancing edge contrast. Consequently, it introduces improvement in the apparent spatial resolution relative to contact imaging[33], in terms of a negative variance equal to $-4\alpha^2$. This variance quantifies the edge-enhancing effect of propagation as the inverse of Gaussian blurring and is obtained by direct calculations in[33]. The forward transfer of the propagation, $T_{prop}(\mathbf{f}) = 1 + \alpha^2(2\pi \mid \mathbf{f} \mid)^2$, relates the spectrum of the transmitted intensity at the sample's exit plane to the propagated intensity; the single-distance Paganin retrieval filter, $H_{pag}$, is the inverse of the propagator[30]. The finite source size introduces geometric blur. The detector PSF subsequently modulates the intensity distribution, also by blurring the image. The observed image therefore reflects the competition between propagation-induced sharpening and system blur. Previous work has shown that this competition can be optimized; there is a specific propagation distance at which the sharpening introduced by propagation maximally counteracts the system PSF blurring, producing the narrowest effective system response. This effect, termed *deblur by defocus*[35], automatically (i.e., in hardware as opposed to "post-processing") achieves spatial deconvolution and phase retrieval[38]. Furthermore, in cases where the system blur is not negligible, it is known that the convolution of a propagated image with the system PSF (Step 1 in the chain) is equivalent to a partial "phase-retrieval" and a reduced regularization parameter, $\alpha'^2 = \gamma' R' \lambda/(4\pi)$, where $\gamma' = \gamma - (\frac{4\pi\sigma_{sys}^2}{2R'\lambda})$, instead of $\alpha^2$, should be used in the Paganin filter[39].

Encoding does not alter the image noise and SNR [37]. Noise is introduced during Step 2, where the measured image is affected by both the source-attributed noise variance, termed self-noise (typically with Gaussian statistics), and the photodetection (shot-noise) variance (typically with Poisson statistics); these are statistically independent and combine linearly in the total noise variance[37]. Photon detection events are spatially distributed by the detector PSF, as such individual photons may contribute to multiple adjacent pixels, introducing correlations in the measured noise, depending on the size of the PSF relative to the pixel size. Because the detector PSF is a linear shift-invariant filter, it smooths both the signal and the noise over the same spatial extent. Consequently, the noise becomes correlated within the width of the detector PSF, while the signal undergoes the corresponding loss of spatial resolution. This correlation reduces the local noise variance and increases the SNR, at the cost of reduced spatial resolution, consistent with the noise-resolution uncertainty (NRU) principle and the invariance of the SNR-to-resolution ratio under linear shift-invariant transformations.

In contrast, source blur arises from the finite size of the X-ray focal spot and acts prior to detection. In the case of cross-spectrally pure radiation with coherent time much shorter than the exposure time, a condition which is true for typical laboratory and synchrotron X-ray imaging, the variance created by self-noise is generally much smaller than the Poisson (shot-noise) contribution and is therefore neglected[36]. This means that photon emission events remain statistically independent and as a result, self-noise does not introduce correlations in the detected shot noise. Consequently, source blur does not provide the same mechanism for variance reduction as detector blur, under the conditions described above.

The conventional Paganin phase-retrieval filter, Step 3, is applied after the noisy image has already been formed; it reduces noise variance by correlating and suppressing high-frequency noise components. This process does not violate the NRU[33], since it also increases the width of the PSF of the imaging system in the same proportion as it decreases the image noise. Therefore, there is no overall change in the spatial resolution because of the combined result of the forward propagation and the retrieval (compensating each other), while the SNR increases during retrieval, leading to the so called "unreasonable" effectiveness of TIE-Hom imaging (TIE-Hom "paradox")[37].

Summarizing, a fundamental distinction between source blur and detector blur has been possibly overlooked. Both source and detector contribute to blur of the image and can be counterbalanced by the apparent edge enhancement due to propagation (Eq. (1)), through *deblur by defocus* by replacing or at least reducing the optimal value of the regularization parameter of the Paganin retrieval[27,39]. However, the detector PSF reduces noise variance per pixel and increase SNR similarly to Paganin filtering (and in general to any filtering), whereas source self-noise does not. The above consideration has an important implication. Only the detector PSF, when it matches the Paganin transfer function over the measurable spatial-frequency band, can perform the role of phase-retrieval directly at the acquisition stage, while also providing a net increase of SNR without loss in resolution, leading to "unreasonable" effectiveness.

For a cone-beam X-ray micro-CT system with a fixed source to detector distance, $R_{tot}$, source to sample distance $R_1$, and sample to detector distance $R_2 = R_{tot} - R_1$, the geometric magnification is $M = R_{tot}/R_1$ . Then, the detector blur, assumed to be approximated by a Gaussian with a detector-plane standard deviation of $\sigma_{det}$, and the source blur, assumed to be approximated by a Gaussian with a source-plane standard deviation of $\sigma_s$, can be mapped into the sample plane by dividing length scales by $M$. The total system blur,

$$\sigma_{sys} = \sqrt{(1-\tfrac{1}{M})^2\sigma_s^2 + (\tfrac{1}{M})^2\sigma_{det}^2} = \sqrt{\sigma_{s_eff}^2 + \sigma_{\det_eff}^2}, \quad (3)$$

can then be calculated at the sample plane, where $\sigma_{s_eff}$ and $\sigma_{det_eff}$ are the source and detector standard deviation at the sample-plane, respectively. Calculating the system's blur Fourier transform:

$$Hsys(\mathbf{f}) = \exp\left(-2\pi^2|\mathbf{f}|^2\left[\left(\tfrac{1}{M}\right)^2\sigma_{det}^2\right]\right)\exp\left(-2\pi^2|\mathbf{f}|^2\left[\left(1-\tfrac{1}{M}\right)^2\sigma_s^2\right]\right), \quad (4)$$

and substituting it into Eq. (2), gives

$$\widehat{\widetilde{I_{\det}}}(\mathbf{f}) = [1 + \left(\frac{R_2\lambda\gamma}{4\pi M}\right)(2\pi \mid \mathbf{f} \mid)^2]\widehat{I_0}(\mathbf{f})$$
$$\exp\left(-2\pi^2|\mathbf{f}|^2\left[\left(1-\tfrac{1}{M}\right)^2\sigma_s^2 + \left(\tfrac{1}{M}\right)^2\sigma_{det}^2\right]\right). \quad (5)$$

This equation, consisting of the propagation term, the transmitted intensity spectrum at the sample exit plane, and the system blur, relates the normalized detected spectrum with the imaging system geometry and characteristics, namely $R_2$, $M$, $\lambda$, $\sigma_s$, $\sigma_{det}$, and the sample's $\gamma$ value. Equation (5) further shows that, in the presence of propagation, the effective spatial resolution of the acquired image is not determined by the system PSF alone. Given that the two-dimensional variance of the (assumed circular) Gaussian system PSF is $\mathrm{Var}[P_{\mathrm{sys}}] = 2\sigma_{\mathrm{sys}}^2$, the negative variance that the propagation contributes to the effective variance of the propagated system response, $-4a^2$, corresponds to a contribution $-2a^2$ to the Gaussian variance parameter $\sigma^2$. The Gaussian-equivalent effective width of the propagated response is

$$\sigma_{eff}^2 = \sigma_{s_eff}^2 + \sigma_{\det_eff}^2 - 2a^2. \quad (6)$$

Thus, whenever propagation contributes non-negligibly to image formation, the effective spatial resolution can be finer than that predicted by the conventional system blur (Eq. (3)). It should be noted here that the phase-dependent effective system width, $\sigma_{eff}$, depends not only on the imaging system and geometry, but also on the sample, through $a^2$.

Following Eq. (5), the retrieved sample's spectrum at $z = 0$ can be estimated by applying a regularized inverse of the propagation operator (i.e., the Paganin filter) to the detected spectrum,

$$\widehat{I_r}(\mathbf{f}) = \widehat{\widetilde{I_{\det}}}(\mathbf{f})\ [1 + \left(\frac{R_2\lambda\gamma}{4\pi M} - \frac{\left(1-\frac{1}{M}\right)^2\sigma_s^2 + \left(\frac{1}{M}\right)^2\sigma_{det}^2}{2}\right)(2\pi \mid \mathbf{f} \mid)^2\ ]^{-1}$$

$$= \widehat{I_0}(\mathbf{f})\exp\left(-2\pi^2|\mathbf{f}|^2\left(1-\frac{1}{M}\right)^2\sigma_s^2\right)\exp\left(-2\pi^2|\mathbf{f}|^2\left(\frac{1}{M}\right)^2\sigma_{det}^2\right)\frac{1+\left(\frac{R_2\lambda\gamma}{4\pi M}\right)(2\pi\mid\mathbf{f}\mid)^2}{1+\left(\frac{R_2\lambda\gamma}{4\pi M}-\frac{\left(1-\frac{1}{M}\right)^2\sigma_s^2+\left(\frac{1}{M}\right)^2\sigma_{det}^2}{2}\right)(2\pi\mid\mathbf{f}\mid)^2}$$

$$= \widehat{I_0}H_sH_{det}T_{prop}H'_{pag} \qquad (7)$$

The reduced regularization parameter of Paganin filter, $H'_{pag}$ in Eq. (7) accounts for the (partial or total) phase retrieval caused by the system blur [39]. Equation (7) shows that for an ideal imaging system with zero blur ($\sigma_s = \sigma_{det} = 0$), $H'_{pag} = H_{pag} = \frac{1}{T_{prop}}$, meaning that the reconstructed spectrum is the sample's exit-surface intensity spectrum, i.e., the Fourier transform of the signal $I_0(\mathbf{r})$.

Micro-CT systems can be classified into two regimes: Regime A, absorption imaging, and Regime B, phase-transfer imaging.

In Regime A, absorption imaging, propagation-induced phase transfer is negligible compared with the blurring imposed by the imaging system. This may arise from a negligible effective propagation distance, $R_2 \sim 0$, a sufficiently small $\gamma$ value of the material at the energies of the X-ray spectrum, or system blur that strongly dominates the propagation-induced sharpening, such that, $\sigma_{sys}^2 \gg 2a^2$. In this case, the propagation term has a negligible effect on the detected spatial-frequency content, and Eq. (7) reduces to

$$\widehat{I_r}(\mathbf{f}) = \widehat{I_0}(\mathbf{f})\exp\left(-2\pi^2|\mathbf{f}|^2\left(1-\frac{1}{M}\right)^2\sigma_s^2\right)\left(-2\pi^2|\mathbf{f}|^2\left(\frac{1}{M}\right)^2\sigma_{det}^2\right). \qquad (8)$$

The image formation is therefore effectively absorption dominated, and the conventional system blur determines the spatial resolution.

In Regime B, phase-transfer imaging, propagation-induced phase transfer is non-negligible relative to the system blur. This requires non-zero propagation distance, $R_2 > 0$, and a non-negligible $\gamma$, together with $2a^2$ non-negligible relative to $\sigma_{sys}^2$. Propagation therefore contributes appreciably to the detected spatial-frequency content and modifies the effective system response, regardless of whether fringes are visible or not. Because practical imaging systems have a finite spatial response, some degree of hardware-induced phase retrieval (HIPR) accompanies phase-transfer imaging in Regime B. Depending on the relative contributions of propagation, source blur, and detector blur, this regime can be further divided into three subcategories: under-HIPR, matched-HIPR, and over-HIPR. In under-HIPR, the hardware blur is insufficient to fully compensate the propagation-induced sharpening. This is the conventional free-space propagation phase-contrast imaging. Residual edge enhancement in the form of fringes may remain (depending on noise and sampling) and additional Paganin (software) retrieval is required. In matched-HIPR, hardware blur approximately matches the Paganin phase-retrieval filtering required to compensate the propagation-induced sharpening. Finally, in over-HIPR, hardware blur exceeds the propagation-induced sharpening. The effective spatial resolution is finer than that predicted by the conventional system blur, and it depends on the sample, Eq. (6). In both matched-HIPR and over-HIPR, additional Paganin retrieval is not required.

These subcategories also provide a framework for optimising imaging systems operating in Regime B. To preserve the phase-induced resolution enhancement, excessive hardware filtering, i.e., over-HIPR, should be avoided. Importantly, the distribution of this filtering between the source and detector blur has different consequences for image noise, as is discussed above. Simulations comparing source-induced and detector-induced hardware retrieval are presented in Supplementary Discussion 8 and Supplementary Figs. 11-12. Consequently, for a fixed total hardware response and fixed photon statistics, allocating most or all of the required filtering to the detector provides the greatest SNR benefit while retaining the phase-enhanced spatial resolution. We refer to this limiting case as detector-induced phase retrieval (DIPR).

In existing micro-CT systems, the condition for DIPR may be inherently met due to the combination of small focal spot, non-zero sample-to-detector distance and detector with non-negligible PSF. In such systems, phase-to-intensity conversion occurs prior to detection, while the detector PSF provides low-pass filtering of the propagated signal. An important difference between DIPR and under-HIPR (requiring conventional Paganin retrieval) is that the former provides an SNR improvement (to the signal before detection) set by the detector response and is therefore fixed for a given imaging system and geometry, whereas the latter provides an SNR improvement (to the signal after acquisition), set by the filter's strength. As a result, for a fixed detector response, acquisition geometry, spectrum and incident statistics, the SNR of the images acquired under DIPR is fixed, rather than retrospectively adjustable. The sample's phase effects instead govern the apparent spatial resolution enhancement due to propagation (pre-detection phase-to-intensity conversion).

We now consider the optimization of the imaging system operating in Regime B for DIPR, i.e., for a resolution-preserving and an SNR-beneficial implementation. A prerequisite for DIPR is that the source blur does not compensate the propagation-induced sharpening before detection. For the Gaussian approximation, this requires sufficient spatial coherence, and is expressed by:

$$a^2 - \frac{{\sigma_{s_eff}}^2}{2} \gg 0. \tag{9}$$

Equation (9), which corresponds to the finite-source coherence criterion for each imaging geometry, ensures that a residual propagation-induced sharpening after finite source blur remains for the detector response to retrieve. Also, in DIPR, the detector MTF approximates the Paganin transfer function. In this case, the detector response (blur) compensates the propagation-induced edge enhancement directly during acquisition, such that $T_{prop}H_{det}\approx 1$. The detected image therefore approximates the exit-surface intensity distribution, with the residual blur governed by source blur,

$$\widehat{I_r}(\mathbf{f}) = \widehat{I_0}(\mathbf{f})\exp\left(-2\pi^2|\mathbf{f}|^2\left(1-\frac{1}{M}\right)^2\sigma_s^2\right), \tag{10}$$

which, in DIPR, is by definition either negligible or small. Equation (10) describes the continuous-domain case; in practice, however, the measured image is additionally subject to discrete detector sampling. Consequently, even when $H_{\mathrm{det}}(\mathbf{f}) \approx H_{\mathrm{Pag}}(\mathbf{f})$, the achievable spatial resolution may remain limited by the finite detector sampling, as higher spatial frequencies are no longer meaningfully represented after sampling. The imaging system may therefore become sampling-limited, such that the measured spatial resolution is coarser than the source-limited prediction of Eq. (10). Throughout this work, we distinguish between the Gaussian-equivalent system blur and the sample-plane sampling pitch[42], and classify an imaging geometry as sampling-limited whenever the former is smaller than the latter. In DIPR, the smallest resolvable feature is governed by the detector sampling (rather than the system blur), and the experimentally measured spatial resolution typically approaches 2 effective (sample-plane) pixels, consistent with the Nyquist-Shannon sampling theorem[43]. This motivates the introduction of frequency-domain metrics evaluated only over the spatial-frequency range that can be reliably transferred by the sampled imaging system.

We define two frequency-domain metrics for DIPR system optimization: phase transfer capability prior to detection and DIPR compatibility. Both metrics are evaluated over the spatial-frequency range in which phase information remains available after finite-source blurring and is meaningfully represented by the sampled detector image $f_1 \leq f \leq f_{\max}$. To determine this range, the function, $G(\mathbf{f}) = H_s(\mathbf{f})\, H_{pix}(\mathbf{f})$, representing the combined attenuation of the propagated phase signal by finite source blur and pixel aperture averaging prior to sampling, is compared with its first sampling replica shifted by the sampling frequency $1/p$, where $p$ is the effective sample-plane pixel size. The intersection of the two curves marks the onset of spectral overlap and defines the upper integration limit, $f_{max}$ (see Supplementary Fig. 8). If no spectral overlap occurs, $f_{max}$ is taken as the first spatial frequency at which the measured detector MTF falls below $10^{-3}$, corresponding to a negligible detector response. The lower integration limit is chosen as $f_l = 0.02 f_{max}$, to reduce the influence of the DC component and slowly varying absorption background.

The first metric quantifies the amount of propagation-induced phase transfer available within the usable spatial-frequency range of the imaging system, considering both the finite source size and the effective propagation distance $R'$. The pre-detector phase-to-intensity transfer is modelled as the product of the projected finite-source MTF and the Fresnel propagation transfer:

$$H_{\phi\to I}^{\text{pre}}(\mathbf{f}) = H_{\text{s}}(\mathbf{f}) \sin(\pi\lambda R'\mathbf{f}^2), \tag{11}$$

where $H_s$ describes the dampening of spatial frequencies due to projected finite source size. We then define a pre-detector phase transfer magnitude:

$$\Psi_{\phi}^{pre} = \sqrt{\frac{\int_{f_1}^{f_{\max}} |H_{\phi\to I}^{\text{pre}}(\mathbf{f})|^2 d\mathbf{f}}{f_{max}-f_1}}. \tag{12}$$

The integral represents the phase-transfer energy over the selected spatial-frequency band. Division by the bandwidth $f_{\max} - f_1$, yields the band-averaged squared phase-transfer magnitude, while the outer square root gives the corresponding root-mean-square phase-transfer magnitude. This normalization ensures that $\Psi_{\phi}^{pre}$ is dimensionless, independent of the specific bandwidth width, and comparable across geometries under a fixed detector frequency band definition. To isolate the degradation of the phase transfer due to finite source size and non-optimal propagation geometry, we normalize this quantity to its coherent-limit maximum, $\Psi_{\phi_max}^{\text{pre}}$, corresponding to a point source and most effective propagation distance $R'_{max}$, which occurs at $R_{tot}/2$ distance for a system with a fixed source-to-detector distance. The quantity:

$$n_{\varphi}(R_1) = \frac{\Psi_{\phi}^{\text{pre}}}{\Psi_{\phi_max}^{\text{pre}}}, \tag{13}$$

which ranges from 0 to 1, represents the fraction of the coherent-limit phase transfer preserved within the system-specific usable frequency band. A value of $n_{\phi} = 1$ indicates that the system achieves the maximum physically attainable phase transfer over the spatial-frequency band supported by the finite source and detector sampling, whereas lower values quantify the loss of usable phase contrast due to finite source size and non-optimal propagation conditions.

The second metric evaluates the DIPR compatibility. It quantifies the similarity between the detector's MTF and the Paganin filter using the Root Mean Square (RMS) match:

$$E_{\text{match}}(r_1) = \sqrt{\frac{\int_{f_1}^{f_{\max}} \left[H_{\text{det}}(\mathbf{f}) - H_{\text{Pag}}(\mathbf{f})\right]^2 d\mathbf{f}}{f_{max} - f_1}}. \tag{14}$$

This is defined as a weighted RMS difference normalized by the measurable detector bandwidth $f_{\max} - f_1$, ensuring this metric is a dimensionless measure of relative spectral shape deviation between the detector response and the Paganin transfer function. Both $H_{det}$ and $H_{pag}$ are normalized to unity at zero frequency.

Finally, it should be noted that the detector PSF is not always approximated by a Gaussian, as is assumed in Eq. (3). Instead, it often exhibits long tails, with its distribution being approximated more closely by a mixed Gaussian-Lorentzian model. In such cases, the conventional Gaussian standard deviation $\sigma$, used to combine independent blur contributions in quadrature, is, strictly, no longer applicable[33]. To provide a spatial resolution measure that remains valid for both Gaussian and non-Gaussian PSFs, we adopt the resolution parameter $\widetilde{\Delta_{sys}}$ proposed by Gureyev et al.,[33], which remains finite for both Gaussian and Lorentzian distributions. Considering a detector described by a mixed Gaussian–Lorentzian model, comprising a Lorentzian component with half-width at half maximum (HWHM) $\gamma_L$ and relative fraction $L$, and a Gaussian component with standard deviation $\sigma_G$ and relative fraction $1 - L$, the generalized system resolution parameter is obtained by combining the projected source and detector contributions in quadrature,

$$\widetilde{\Delta_{sys}} = \sqrt{\left(2\sqrt{\pi}\sigma_s\right)^2 + \left[(1 - L)2\sqrt{\pi}\sigma_G + L2\pi\gamma_L\right]^2}. \tag{15}$$

For consistency with conventional Gaussian resolution metrics, this generalized resolution parameter is then converted to an equivalent Gaussian FWHM,

$$FWHM_{sys} \approx \widetilde{\Delta_{sys}} \frac{2\sqrt{2\ln 2}}{2\sqrt{\pi}}, \tag{16}$$

where the conversion follows directly from the analytical relationship between the generalized resolution parameter $\widetilde{\Delta_{sys}}$ and the FWHM of a Gaussian PSF[33]. For the avoidance of doubt, the transformation in Eq. (16) does not assume that the system blur is Gaussian, it just provides an equivalent Gaussian width.

**Experimental scans**

Two micro-CT systems were used in this work: a custom-built system and a commercial Nikon scanner. A schematic representation of our custom system is shown in Fig. 3 (a).

A Hamamatsu microfocus source with a Tungsten anode (model L12161-07) was used with a tube voltage of 60 kVp and a tube current of 166 µA. Two focus modes, small and large, were utilized for the acquired scans, leading to a different size of the (to good approximation Gaussian) focal spot. The FWHM of the source with the small and large focus mode was estimated to be 12 ± 3 µm and 63 ± 7 µm, respectively. Prior to any acquisitions, the source was allowed to stabilize for 2 hours. A Hamamatsu CMOS-based flat panel (model C9732DK) with 2368 (h) × 2340 (v) 50 × 50 µm$^2$ pixels was used. Its PSF was measured to be approximated by a mixed Gaussian–Lorentzian model, comprising 67.9% of a Lorentzian with 50 µm HWHM and 32.1% of a Gaussian with 130 µm standard deviation (see Supplementary Fig. 9). The source-to-detector distance, $R_{tot}$, was 919 mm. The sample stage consisted of a Newport linear stage (model M-ILS150BPP) for the positioning of the sample along the X-ray beam propagation (i.e., along the $z$ axis), a SmarAct rotation stage (model SR-5014-S) for the angular rotation, and a SmarAct translation stage (model SLC-24150-W-S) for the positioning of the sample along the $x$ axis.

Four scans, Scan 1 – Scan 4, were acquired, the experimental details of which are summarized in Table I. The magnification between scans varied from 1.5 ($R_1$ = 605 mm) to 20.4 ($R_1$ = 45 mm), in order to vary the source's and detector's contributions to the overall system blur[40,44], thereby allowing transition from Regime A to Regime B. 720 projections were acquired from 0° to 359.5° with a 0.5° step and a 0.5 s exposure time for Scan 1 and 2 s exposure time for Scan 2 – Scan 4. For each scan, 10 dark and 10 flat images were acquired prior to the sample projection images for dark and flat field correction.

The Nikon scanner was an XT H 225 with a Tungsten anode. A tube voltage of 60 kVp and a tube current of 117 µA were set to achieve the smallest focal spot size; this was approximated by a Gaussian with 17 µm FWHM (see Supplementary Fig. 10). A flat panel detector with 2028 (h) × 2028 (v) 200 × 200 µm$^2$ pixels was used. Its PSF was measured to be approximated by a mixed Gaussian–Lorentzian model, comprising 40% of a Lorentzian with 120 µm HWHM and 60% of a Gaussian with 286 µm standard deviation (see Supplementary Fig. 9). The source-to-detector distance, $R_{tot}$, was 911 mm. Two scans, Scan I and Scan II, were acquired; experimental details are summarized in Table I. The magnification was 7, corresponding to $R_1$ = 130 mm and Regime B – DIPR. 1000 projections were acquired from 0° to 359.64° with a 0.36° step and a 2 s exposure time. Scan I was acquired without dithering; dithering is a sampling strategy in which multiple acquisitions are recorded with sub-pixel relative shifts between the sample and detector. Scan II employed half-pixel dithering, consisting of two identical CT acquisitions, with the second acquired after translating the sample by half an effective detector pixel (14.3 μm) along the $x$ direction, to overcome the relatively coarse sample-plane sampling. The two projection datasets were interleaved prior to reconstruction, yielding a twofold increase in sampling density without altering the detector aperture.

The scanned samples were a tissue-engineered esophageal scaffold[45] and a 1 mm diameter PTFE wire.

**Theoretical calculations and simulations**

To evaluate the proposed framework, the theoretical metrics, namely finite-source coherence criterion, Eq. (9), normalized pre-detector phase-transfer metric, Eq. (13), and detector compatibility metric, Eq. (14), were calculated as functions of source-to-sample distance, $R_1$, for our custom system and the Nikon scanner. The calculations were performed for a 1 mm diameter PTFE wire using the mean energy of the corresponding experimental X-ray spectrum (Table I).

Numerical simulations were performed using a 1D wave optics-based model[46] along the detector $x$-direction for each projection angle. The model is based on the Fresnel–Kirchhoff diffraction theory within the Fresnel approximation. A monoenergetic X-ray beam of 17.6 keV was considered for Scan 1 – Scan 4 and 25.0 keV for Scan I and Scan II. These were selected to reproduce the experimentally measured attenuation of the 1 mm diameter PTFE wire for the corresponding imaging systems, which was also the sample implemented in the simulations, modelled as a complex transmission function. Free-space propagation was simulated using the Fresnel transfer function, followed by finite-source blurring and pixel aperture averaging at the sample-plane effective pixel size. To account for the detector response, each simulated 1D projection was replicated along the $y$ direction to form a 2D projection image and was then convolved with the measured 2D detector PSF, modelled as a mixed Gaussian–Lorentzian distribution. Poisson noise was added before detector convolution. The noise level in the simulated image was estimated from the experimental dark corrected flat-field images for Scan 2 and subsequently scaled for the remaining imaging geometries according to the corresponding effective detector pixel size (i.e., accounting for changes in geometric magnification and exposure time). All functions were sampled along the horizontal directions with a 50 nm step size.

Further simulations examining the practical implementation of matched-HIPR are provided in Supplementary Discussion 8 and Supplementary Figs. 11-12.

**Data analysis**

Software-based phase retrieval for both the experimental and simulated scans was performed using the single-distance algorithm developed by Paganin et al. [30]. CT reconstruction of the experimentally acquired raw (edge-enhanced attenuation) and phase-retrieved images was carried out using a GPU-accelerated implementation of the Feldkamp–Davis–Kress (FDK) cone-beam algorithm[47] provided in the ASTRA toolbox[48,49]. CT reconstruction of the simulated raw and phase-retrieved datasets was performed using filtered back-projection.

The data analysis also included quantifying the relative fringe amplitude at the edges of the PTFE wire in the projection images and determining the SNR and spatial resolution in the reconstructed PTFE wire volumes. The fringe amplitude measurements were obtained after normalizing the intensity profiles to the mean background intensity. Multiple profiles were extracted from both the right and left wire edges, over 250 μm of wire length for the experimental images. Mean values and standard deviations were then calculated. The SNR was computed as the ratio of the mean signal intensity within the wire to the standard deviation of the background signal. Spatial resolution was estimated from the reconstructed PTFE wire edges, by extracting radial edge profiles extending from the centre of the wire in transverse CT slices. Error functions were fitted to the edge profiles, and their derivatives were computed to obtain the corresponding line-spread functions (LSFs). The full width at half maximum (FWHM) of the LSFs was then extracted. For the simulated images, SNR and spatial resolution values were obtained across 10 CT slices, and their mean and standard deviation were reported. The projected source blur, detector blur, total system blur and Paganin transfer function were calculated and compared in both the spatial and frequency domains.

Table I. Details of the four experimental scans with our custom system, Scan 1 – Scan 4, and the two experimental scans with the Nikon scanner, Scan I and Scan II. The FWHM of detector blur and total system blur correspond to an equivalent Gaussian FWHM, see Eq. (16).

| | **Scan 1** | **Scan 2** | **Scan 3** | **Scan 4** | **Scan I** | **Scan II** |
|---|---|---|---|---|---|---|
| $R_{tot}$ (mm) | 919 | 919 | 919 | 919 | 911 | 911 |
| $R1$ (mm) | 45 | 305 | 500 | 605 | 130 | 130 |
| Mean X-ray energy (keV) | 17.6 | 17.6 | 17.6 | 17.6 | 25.0 | 25.0 |
| Magnification | 20.4 | 3.0 | 1.8 | 1.5 | 7.0 | 7.0 |
| Detector pixel size (μm) | 50 | 50 | 50 | 50 | 200 | 200 |
| Sample-plane pixel size (μm) | 2.4 | 16.6 | 27.2 | 32.9 | 28.5 | 14.3 |
| Exposure time (s) | 0.5 | 2 | 2 | 2 | 2 | 2 |
| Detector blur FWHM (μm) | 8.9 | 60.9 | 99.8 | 120.8 | 53.0 | 53.0 |
| Source blur FWHM (μm) | 11.4 | 8.0 | 5.5 | 4.1 | 14.6 | 14.6 |
| System blur FWHM (μm) | 14.5 | 61.4 | 99.9 | 120.8 | 55 | 55 |

## Data availability

The data that support the findings of this study are available from the corresponding author upon reasonable request.

## Acknowledgements

This work was funded by EPSRC (Grant EP/T005408/1). AO is supported by the Royal Academy of Engineering under their Chairs in Emerging Technologies scheme (Grant CiET1819/2/78). CKH is supported by the Royal Academy of Engineering under the Research Fellowship Scheme (Grant RF201617\16\27).